\documentclass[reprint, amsmath,amssymb,aps,]{revtex4-1}
\usepackage{multirow}
\usepackage{mathtools}
\usepackage{bigstrut}
\usepackage[colorlinks]{hyperref}
\RequirePackage{mathptmx}      
\RequirePackage{flushend}
\usepackage{ulem} 

\def\im		{\ensuremath{{\rm Im}}}
\def\re		{\ensuremath{{\rm Re}}}
\def\xst	{\ensuremath{\sigma_{\rm tot}}}
\def\ba		{\begin{eqnarray}}
\def\ea		{\end{eqnarray}}
\def\no		{\nonumber}
\def\lt		{\ensuremath{\left(}}
\def\rt		{\ensuremath{\right)}}
\def\lq		{\ensuremath{\left[}}
\def\rq		{\ensuremath{\right]}}
\def\chin	{\ensuremath{\overline{\chi}^2}}
\def\hh		{\ensuremath{\hspace{4mm}}}
\def\ds		{\ensuremath{\displaystyle}}

\def\ro0{$\rho_{j}(t=0)$}

\begin{document}

\preprint{LNF/2018}
\title{Analysis and implications of precision near-forward TOTEM data}

\author
       {Simone Pacetti} \email{simone.pacetti@pg.infn.it} 
      \author{Yogendra Srivastava} \email{yogendra.srivastava@gmail.com}
         \altaffiliation{Also at Physics Department, Northeastern University, Boston, Massachusetts 02115, USA}
        \affiliation{Dipartimento di Fisica e Geologia, Universit\`a degli Studi di Perugia and \\ 
        INFN Sezione di Perugia, 06123 Perugia, Italy}
        \author{Giulia Pancheri}\email{giulia.pancheri@lnf.infn.it}
                \altaffiliation{Also at Department of Physics, Massachusetts Institute of Technology 
                Center for Theoretical Physics and Laboratory for Nuclear Science, Cambridge, Massachusetts 02139, USA}
                \affiliation{Laboratori Nazionali dell'INFN di Frascati, 00044 Frascati, Italy}

\date{\today}

\begin{abstract}
Very precise data on elastic proton-proton scattering at $\sqrt{s}=7$, $8$ and $13$ TeV have been obtained by the TOTEM group at LHC in the near-forward region (momentum transfers down to $|t| = 6 \times 10^{-4}\ {\rm GeV}^2$ at $\sqrt{s}~=~8$~TeV and to $|t| = 8 \times 10^{-4}\ {\rm GeV}^2$ at $\sqrt{s}=13$ TeV). The Coulomb-nuclear interference has been measured with sufficient accuracy for TOTEM to establish the falloff of the $\rho$ parameter with increasing energy. The predictions from a previously studied model are shown to be in good agreement with the data and thus allow us to draw rather firm conclusions about the structure of the near-forward nuclear amplitude. We point out that due to a zero in the real part of the nuclear amplitude occurring at a very small momentum transfer--that can migrate into the Coulomb-nuclear interference (CNI) region at higher energies--much care is needed in extracting the numerical value of $\rho$ for such energies. Thus, the true value of $\rho$ would be higher than the TOTEM value for $\rho$ found under the hypothesis that the real part of the elastic nuclear amplitude is devoid of such a zero in the CNI region. 
\end{abstract}
\pacs{Valid PACS appear here}

\maketitle

\section{Introduction \label{intro}}
The TOTEM group has produced a remarkably precise determination of the proton-proton elastic nuclear amplitude at LHC energies~\cite{TOTEM:2011,TOTEM:2016, TOTEM1:2017, TOTEM2:2017}. In particular, through the Coulomb-nuclear interference (CNI) at very small momentum transfer, TOTEM has reported direct measurements of the $\rho$ parameter (that is, the ratio of the real to the imaginary part of the nuclear amplitude in the forward direction at $\sqrt{s} = 8$ and 13~TeV). Thus, we now have data for the modulus (through the elastic differential cross section) and the phase (through CNI) of the near-forward nuclear amplitude.

The energy dependence of the $\rho$ parameter
  has been extensively studied both experimentally and theoretically for over five decades.  Spanning an energy range of more than 3 orders of magnitude, experiments performed with the CERN intersecting storage rings up those at  the Large Hadron Collider (LHC), show the parameter to rise, with $\rho_{ {\bar p}p}$ apparently stabilizing its value around Tevatron energies. Asymptotically, the parameter was predicted a long time ago by Khuri and Kinoshita (KK) in Ref.~\cite{khuri} to  decrease as  $(\pi/\ln s)$, both for $pp$ and ${\bar p}p$. The first TOTEM results at LHC energy $\sqrt{s}=7$~TeV were noncommittal as to whether  the predicted asymptotic decrease had started. By now, the reported TOTEM results at $\sqrt{s}=8$ and 13~TeV point to a decrease, apparently
  even faster than the original KK prediction, as clearly seen from the compilation shown by  Fig.~15 in  Ref.~\cite{TOTEM2:2017}. The question as to  whether this is the  epiphany of a new phenomenon or, the high energy manifestation of nonleading contributions to the $pp$ elastic cross section, such as the odderon~\cite{TOTEM1:2017,Nicolescu:1975}, is presently under discussion. 

In the present paper, we shall analyze  the TOTEM results for the CNI region at $\sqrt{s}=8$ and 13~TeV through 
 a recently studied model~\cite{Fagundes:2013,Pancheri:2017}, which is a modified version of a model originally proposed by Barger and Phillips (BP)~\cite{BP}. The BP amplitude was proposed as an "independent" model to highlight the main characteristics of the elastic diffractive cross section, namely the forward peak, the  dip-bump structure arising from  the zero of the imaginary part of the amplitude in $pp$, the slower, powerlike decrease at large $-t$. In the original  model, the parameter $\rho$ is negative, in agreement with data on $pp$ scattering at the time. However, as the energy increases, besides the well-known "diffraction dip" due to a zero in the absorptive part, the amplitude must develop a zero in the real part of the nuclear amplitude for the parameter $\rho$ to change sign. If, at LHC energies, the real part vanishes in the CNI region, the extraction of the $\rho$ parameter gets complicated. \\
 A zero in the real part of the elastic amplitude has been studied previously by Martin~\cite{Martin:1997} and more extensively for the LHC data at $\sqrt{s}=7$ and 8~TeV by Kohara {\it et al.} in Ref.~\cite{Kohara:2017}, who find a positive evidence for this Martin zero. In fact, Martin~\cite{Martin:1997} proved that if at infinite energies, (i) the total cross section tends to infinity and (ii) the differential elastic cross section tends to zero for large $-t$ values, then the real part of the even-signature amplitude  must change sign near $t=0$.  Such a circumstance  necessarily questions  the assumption, underlying the  present  TOTEM analysis of the  CNI data,  that the real part of the nuclear amplitude is a constant. At LHC energies, data show  the total cross section to rise continuously and the differential cross section at a fast falloff at high $-t$ values, satisfying both of Martin's assumptions. Martin's theorem then implies that, if the real part of the amplitude is positive at $t=0$, due to the existence of a zero  in the CNI region, the value of $\rho$ obtained from an analysis assuming the real part of the nuclear amplitude to be a constant (near $t=0$) would be lower than its true value.
\\

The paper is organized as follows. In Sec.~\ref{geo} the presence of a zero in the real part of the forward amplitude is discussed in the general case of geometrical scaling. In Sec.~\ref{BP}, we discuss various prescriptions for obtaining the real and imaginary parts of the modified BP nuclear amplitude based on our phenomenological expressions for the parameters deduced in Ref.~\cite{Fagundes:2013}. In Secs.~\ref{differential} and~\ref{sec:noi}, after a brief description of the parametrization for the nuclear amplitude used by TOTEM, we compare the results from our model with the TOTEM data at $\sqrt{s}=8$ and 13~TeV in the CNI region. Relative residuals and $\chi^2$'s are defined and discussed in detail. In Sec.~\ref{complex}, we exhibit the phases and both real and imaginary parts of all four of our nuclear amplitudes in order to understand better qualitative differences between them and to delineate further the problems associated with finding the {\it correct} value of $\rho$ (at $t=0$). We shall see that our {\it rotated} BP amplitudes move the zero in the real part of the elastic amplitude nearer to the CNI at higher energies. Thus, at $\sqrt{s}=13$~TeV, the effect is more pronounced (reducing {\it true} $\rho\simeq 0.13$ to an apparent $t$-averaged $\overline{\rho}\simeq 0.10$; see below), than at $\sqrt{s}=7$ and 8~TeV (as discussed in Ref.~\cite{Kohara:2017}).   

As the real part has a zero in the CNI region, in Sec.~\ref{meanrho} we define mean values of $\rho(s,t)$ in various $t$ intervals to compare our results with those from TOTEM that assumes $\rho$ to be a constant in that $t$ interval. Quite good agreement is reached with the TOTEM values. A compendium of all our results for $\rho$ and [$\sigma_{\rm tot}$, $\sigma_{\rm el}$] is presented and compared with TOTEM data in Tables~\ref{tab:compendiumrho} and~\ref{tab:compendium}. In Sec.~\ref{odd}, we examine the question whether TOTEM data in the CNI region require QCD odderon contributions. In the concluding Sec.~\ref{conc}, we present our conclusions deduced from the model.
\section{\bf Near-forward zero in $\re \lt \mathcal{A}(s,t)\rt$ from Geometric scaling \label{geo}}
In this section we discuss further evidence, arising from geometrical scaling,  for the real part of the nuclear amplitude exhibiting a zero near the forward direction.
\\
If there is a domain in $T=-t$ within which geometric scaling were valid, then Martin~\cite{Martin:1973} showed that 
\begin{eqnarray}
\label{g1}
\re \lt\mathcal{A}(s,T)\rt &=& \rho \frac{d}{dT} \lq T \im \lt\mathcal{A}(s,T)\rt\rq 
\nonumber\\&=&
 \rho \im \lt \mathcal{A}(s,T)\rt + T \frac{d}{dT}\im\lt \mathcal{A}(s,T)\rt\,.
\end{eqnarray}
Since the imaginary part is large and positive at $T = 0$ and decreases to zero at $T =T_I$, then employing Eq.~(\ref{g1}), 
we have:  
\begin{eqnarray}
\label{g2}
\int_0^{T_I} \re \lt\mathcal{A}(s,T)\rt dT = \rho\Big[ T \im \lt\mathcal{A}(s,T)\rt\Big]_{T=0}^{T=T_I} = 0\,.
\end{eqnarray}
Since $\im \lt\mathcal{A}(s,T=0)\rt>0$, satisfaction of Eq.~(\ref{g2}) necessarily implies that the real part of the amplitude must change sign--at least once--somewhere between $T=0$ and $T=T_I$. That is,
\begin{equation}
\label{g3}
\re\lt \mathcal{A}(s,T=T_R)\rt = 0\,;\hspace{3mm} {\rm for}\,\,\ 0< T_R < T_I.
\end{equation}
Geometric scaling is presumably a good approximation for small $T$. What the above analysis shows is that it can be valid up to $T_I$ {\it only if the real part vanishes prior} to the value of $T$ where the imaginary part does.

Actually, one can prove a stronger result: if geometric scaling holds, then between any two consecutive zeroes of the imaginary part, there must be at least one zero of the real part. The trivial proof goes as follows, let $\left\{T_{I,n}\right\}_n$ be a set of zeros for the imaginary part of the amplitude, i.e.,
\begin{eqnarray}
\im\lt\mathcal{A}(s,T_{I,n})\rt = 0\,,\hspace{5mm}  n=1,2,\ldots\,,
\nonumber
\end{eqnarray}
this implies  
\begin{eqnarray}
 \label{g4}
 \int_{T_{I,n}}^{T_{I,n+1}} \re\lt \mathcal{A}(s,T) \rt= 0\,,\hspace{5mm}  n=1,2,\ldots\,,
\end{eqnarray}  
by virtue of Eqs.~\eqref{g1} and~\eqref{g4}.  For the integral in Eq.~(\ref{g4}) to vanish, clearly the real part of the amplitude must change sign at least once, thus proving the theorem.\\
The relevance of the above to the extraction of $\rho$ at LHC energies (and beyond) is obvious. Since the imaginary part of the amplitude at LHC energies has a zero at a rather small value of $T$, i.e., $T_I =(0.45\div 0.55)$ GeV$^2$, geometric scaling tells us the real part has a zero at an even smaller value of $T$. We shall confirm such a behavior within a specific $s\leftrightarrow u$ symmetrized version of the model elastic amplitude presented in Ref.~\cite{Fagundes:2013}.  This will further  support the assertion that care must be exercised in extracting a value for $\rho$ from small $T$ data due to a zero in the near-forward real part of the nuclear amplitude.
\section{Modified Barger and Phillips amplitude \label{BP}}
Let the elastic amplitude $\mathcal{A}(s,t)$ be defined through its relation with the total cross section as
\begin{equation}
\label{B1}
\xst(s) = 4 \sqrt{\pi}\, \im\left( \mathcal{A}(s,0)\right)\,, 
\end{equation}
so that the elastic differential cross section reads
\ba
\label{B2}
\frac{d\sigma}{dt}(s,t) =  |\mathcal{A}(s,t)|^2\ ,
\ea
(all particle masses are being ignored). The amplitude of our modified BP model is~\cite{Fagundes:2013} 
\begin{equation}
\label{B3}
\mathcal{A}(s,t) = i \left[  F^2(t) \sqrt{A} \, e^{Bt/2} + e^{i\phi} \sqrt{C} \,e^{Dt/2}\right]\,.
\end{equation}
In Fagundes {\it et al.}~\cite{Fagundes:2013}, $pp$ data for the elastic differential cross section in the energy range $\sqrt{s}=(24\div7000)$~GeV had been fitted with the above model. Using values from these fits, values of the parameters in Eq.~\eqref{B3}, based on asymptotic theorems and   sum rules \cite{Froissart:1961,Martin:1963,Pancheri:2005} had been put forward, as shown in the following.
\begin{itemize}\item The $s$ dependence of $A$ was chosen so as to saturate the Froissart bound
\begin{eqnarray}
\label{B4}
4 \sqrt{\pi\, A(s)} =
  \left(  0.398\ L^2(s)  - 3.80\ L(s) + 47.8\right)\,{\rm mb}\,,
  \ea
  where
  \ba
L(s) = \ln(s/s_0)\,,\hspace{4mm} s_0 \equiv 1\, {\rm GeV}^2\,.
\label{eq:Ls}
\ea
\item The proton form factor is defined in terms of the standard dipole
\begin{equation}
\label{B6}
F(t) = \frac{1}{(1 - t/t_0(s))^2}\,.
\end{equation}
\item While the phase $\phi$, introduced in Eq.~\eqref{B3}, is very slowly varying with $s$, the pole $t_0(s)$ of the form factor has a not negligible dependence on the energy, as it is shown in Fig.~3 of Ref.~\cite{Fagundes:2013}. However, for large values of $s$, it appears  to approach the usual value of 0.71 GeV$^2$. Hence, for high energies, we had frozen both phase and pole position as
\begin{equation}
\label{B7}
\phi = 2.74\,,\hspace{4mm} 
t_0 = 0.71\ {\rm GeV}^2\,.
\end{equation}
\item 
Based on two asymptotic sum rules~\cite{Pancheri:2005}, which were shown to be almost saturated at $ \sqrt{s}=7 $~TeV~\cite{Fagundes:2013}, the two slope parameters in the model were parametrized as
\ba
\label{B8}
\begin{array}{rcl}
B(s) &=& \left( 0.028\ L^2(s) - 0.230 \right)\ {\rm GeV}^{-2}\,;\\
&&\\
D(s) &=& 
\left( 0.29\ L(s) - 0.41\right)\ {\rm GeV}^{-2}\,. \\
\end{array}
\ea
\item The $s$ dependence of the $C$ term is more complicated due to its large variation from low to high energies. Phenomenologically, we had chosen for an asymptotically constant $C$, the following form
\ba
\label{B5}
4 \sqrt{\pi\, C(s)} = \frac{9.60 - 1.80\ L(s) + 0.01\ L^3(s)}{1.200 + 0.001\ 
L^3(s)} {\rm mb}\,.
\ea
\end{itemize}
Before discussing our application to the recent $\sqrt{s}=13$~TeV TOTEM data, we stress that the asymptotic behavior of the slope $B(s)$ chosen above, differs from the usual $\ln(s)$ behavior expected from Regge or Pomeron pole  trajectories. Namely, in this empirical model, the recent TOTEM observation that the slope in the forward region increases faster than a logarithm, is not a surprise. Our observation is also in agreement with an earlier study  by Shegelsky and Ryskin~\cite{Schegelsky:2012}. For a very comprehensive study of the slope--covering different regions of $t$--see Okorokov~\cite{Okorokov:2015}. Our choice for $B(s)$ in Eq.~\eqref{B8} is consistent with solution (d) of Ref.~\cite{Okorokov:2015}. A recent analysis by Jenkovszky {\it et al.}, in the context of the dipole model, also confirms the acceleration of the slope with energy~\cite{Jenkovszky:2018}.

 To study the CNI region, in light of Martin's observation mentioned in the Introduction~\cite{Martin:1997}, we  first  notice that neither the original BP amplitude nor the amplitude of Ref.~\cite{Fagundes:2013} were $s\leftrightarrow u$ symmetric. In this paper we shall consider modifications of our proposed amplitude, such that the resulting amplitude $\mathcal{A}(s,t)$ would be invariant under $s\leftrightarrow u$, namely we study the even-signature component of the above model.

Let us first consider the transformation rule $\{s \to se^{-i\pi/2}\}$, valid for a $C=+1$ amplitude, as discussed thoroughly by Block in Sec.~10.3 etc. of his review~\cite{Block:2006}. For the present parametrizations, this implies the substitution  $L(s) \to [L(s) - i\pi/2]$ in $A(s)$ and $C(s)$. For example, it gives for the contribution of the $A$ term into the real and imaginary parts of $\mathcal{A}(s,t)$ to be
\ba
\label{B9}
\im \left( 4i \sqrt{\pi  \,A(s)}\right) &=& \left( 
0.398\ L^2(s)  - 3.8\ L(s) + 47.8\right)\,{\rm mb}\,;\no\\
&&\\
\re \left( 4i \sqrt{\pi\,  A(s)}\right) &=& \left(0.398\ \pi\ L(s) - 1.9\ \pi\right)\,{\rm mb}\no\\
\ea
so that their ratio
\ba
\!\!\!\!\!\hat{\rho}(s) &\equiv& \frac{\re\left( i \sqrt{  A(s)}\right) }{\im \left(i \sqrt{ A(s)}\right)}
\lesssim \frac{\pi}{L(s)},
\label{eq:rhohat}
\ea
similarly for $\sqrt{C(s)}$. Of course, such a choice leaves the slope parameters $B(s)$ and $D(s)$ {\it unrotated}.

One may entertain the possibility that the slope parameters $B(s)$and $D(s)$ also get rotated using the above rule $\{s \to se^{-i\pi/2}\}$.  Namely, for a $C = +1$ amplitude, what is really required is to preserve the symmetry  $s\leftrightarrow u$ at fixed $t$ of the entire amplitude~\cite{Martin:1997}. For large $s$ and $u$ at fixed $-t\ll 0$ (neglecting the masses), this reduces to setting $u \to -s = s e^{-i\pi}$. 

In the following, we have considered all four possibilities with the following four sets of nuclear amplitudes:
\begin{eqnarray}
\label{B10}
\mathcal{A}_0 (s,t)&\!\!\!\!\!:&\mbox{no rotation at all}\,;\nonumber\\
\mathcal{A}_1 (s,t)&\!\!\!\!\!:& \mbox{only $A$ and $C$ rotated}
\no\\
&&\mbox{no rotation of the phases $B$ and $D$}\,;\\
\mathcal{A}_2 (s,t) &=& \mathcal{A} (se^{-i\pi/2},t)\,:\ \mbox{complete rotation}\,;\nonumber\\
\mathcal{A}_3 (s,t) &=& \frac{1}{2} \left[\mathcal{A} (s,t) + \mathcal{A} \left(s e^{-i\pi},t\right) \right]\,.\no
\end{eqnarray}
As we shall see, both the real and imaginary parts of the nuclear amplitude are practically identical for $\mathcal{A}_2$ and $\mathcal{A}_3$, but they are substantially different from $\mathcal{A}_0$ and $\mathcal{A}_1$. Moreover, the predictions from the amplitudes $\mathcal{A}_2$ and $\mathcal{A}_3$ are in remarkable accord with the TOTEM data in the CNI region whereas those from $\mathcal{A}_0$ and $\mathcal{A}_1$ are decidedly inferior. In the following Sec.~\ref{differential}, we shall use the amplitudes and parameters of Eqs.~\eqref{B4}-\eqref{B8} for the four nuclear amplitudes of Eq.~\eqref{B10} to make predictions and compare them with the TOTEM data in the CNI region at the two representative energies $\sqrt{s}=8$ and  13~TeV.
\section{Coulomb-nuclear interference\\ at $\sqrt{s}=8$ and 13~TeV \label{differential}}
As stated earlier, the CNI has been measured with great precision by the TOTEM group at $\sqrt{s}=8$ and 13~TeV. Numerical values of the elastic differential cross section in the momentum transfer regions $6\times 10^{-4}< |t| < 0.1961\ {\rm GeV}^2$ for $\sqrt{s}\equiv\sqrt{s_8}=8$~TeV and $8\times 10^{-4}< |t| < 0.202452\ {\rm GeV}^2$ for $\sqrt{s}\equiv\sqrt{s_{13}}=13$~TeV are presented in Tables~3 of Ref.~\cite{TOTEM:2016} and Ref.~\cite{TOTEM2:2017}, respectively. 
 
To isolate CNI, TOTEM shows the $\sqrt{s}=13$~TeV data in the following way~\cite{TOTEM2:2017}.
\begin{enumerate}
	\item Figure~13 of Ref.~\cite{TOTEM2:2017} shows the cross section data with momentum transfer up to $|t| \leq 0.15\ {\rm GeV}^2$.
	\item Figure~14 of Ref.~\cite{TOTEM2:2017} shows cross section data up to $|t| \leq 0.07\ {\rm GeV}^2$.
	\item In both figures, along with data is also plotted the fractional quantity 
	\ba
\label{13.1}
\mathcal{X}(s,t) &=& \frac{\lt d\sigma/dt\rt_{\rm data} - {\rm Ref}_4\lt s,t\rt}{{\rm Ref}_{4}\lt s,t\rt}\,.
\ea
The reference value Ref$_4(s,t)$ at $\sqrt{s}=13$~TeV is defined as 
\ba
{\rm Ref}_4(s_{13},t) &=& 633\ e^{\frac{20.4 \,t}{{\rm GeV}^2}} \,{\rm mb\,\, GeV}^{-2} + \lt\frac{d\sigma}{dt}\rt_C \,.
\label{eq:ref4-13}
\ea
with   the Coulomb cross section,
\ba
\lt\frac{d\sigma}{dt}\rt_C &=& \left|\mathcal{A}^C(s,t)\right|^2\,,
\no
\ea
 defined as 
\ba
\mathcal{A}^C(s,t)=
\frac{2\sqrt{\pi}\,\alpha}{t}\,F^2(t)
\,,
\label{eq:A^C}
\ea 
\end{enumerate} 
The above is the reference nuclear amplitude for  $|t| \le 0.2$~GeV$^2$.
To investigate the CNI, TOTEM~\cite{TOTEM2:2017} uses the following parametrization for the nuclear amplitude, called $\mathcal{A}^N$,
\begin{eqnarray}
\label{13.2}
\mathcal{A}^N(s,t) &=& \left|\mathcal{A}^N(s,t)\right| e^{i\Phi(s,t)}\,;\nonumber\\
\left|\mathcal{A}^N(s,t)\right| &=& \sqrt{a}\, \exp\lt{\frac{1}{2}\sum_{n=1}^{N_b} b_n t^n} \rt\,;\\
\Phi(s,t) &=& \frac{\pi}{2} - \tan^{-1} \lq\rho(s,t)\rq = [{\rm Constant}]\,.\no
\end{eqnarray}
It depends on the set of $N_b+1$ parameters $\left\{a,b_1,b_2,\ldots ,b_{N_b}\right\}$. In particular, $b_1$ is the ``large'' diffraction slope and $b_{2,3}$ are supposed to account for minor fluctuations in the low-$|t|$ data. It is interesting to note that, in Ref.~\cite{TOTEM2:2017}, for the data  covering a smaller $t$ interval, $|t|_{\rm max} = 0.07\ {\rm GeV}^2$,  for $N_b = 1$, 2, 3, the $\chi^2$ per degree of freedom, \chin, have roughly the same value: $\chin=0.7$, 0.6, 0.6, respectively; whereas data that cover a larger $t$ interval, $|t|_{\rm max} = 0.15\ {\rm GeV}^2$, the fit with just one term $N_b =1$ has a much larger $\chin=2.6$ compared to $\chin=1.0$ for $N_b=2$ and $\chin=0.9$ for $N_b=3$. In fact, as the authors of Ref.~\cite{TOTEM2:2017} note themselves, the quality of fit is bad and no values for $\rho$ are displayed for $N_b=1$ and $|t|_{\rm max} = 0.15\ {\rm GeV}^2$. Notice that  their chosen parametrization of the nuclear amplitude leaves no room for the real part of the nuclear amplitude to vanish in the CNI region. It cannot be excluded that this assumption is responsible for the poor fit in the case $N_b=1$, since, as  Martin's theorem indicates, the existence of a zero  is highly likely. \\
To investigate the existence of a zero, in the following Sec.~\ref{sec:noi}, we shall follow  the steps of the TOTEM analysis but using our rotated asymptotic model in lieu of their parametrization, which excludes {\it a priori} the  possibility of  $\rho(s,t)$ to change sign in the CNI region. 
\section{Differential cross section and residuals at $\sqrt{s}=8$ and 13~TeV\label{sec:noi}}
\begin{figure}[h!]
\begin{center}
\includegraphics[width=\columnwidth]{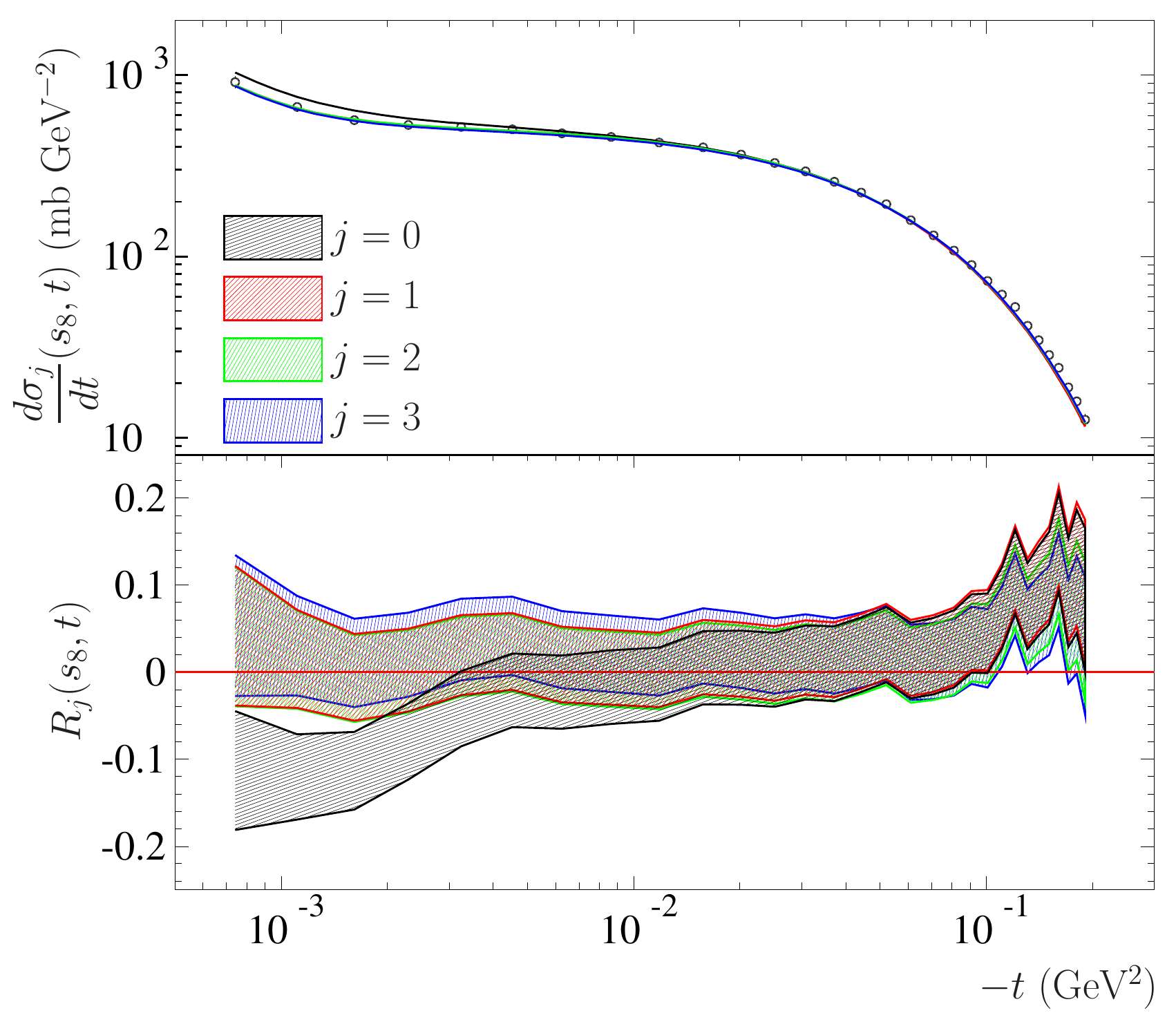}
\caption{\label{fig:diff-xs8}(Upper) Data on the differential cross section at $\sqrt{s}=8$~TeV and superimposed the predictions corresponding to the parametrizations of the nuclear amplitude given in Eq.~(\ref{B10}). (Lower) Residuals as defined in Eq.~(\ref{eq:residuals}).}	\end{center}
	\end{figure}

\begin{figure}[h!]
\begin{center}
\includegraphics[width=\columnwidth]{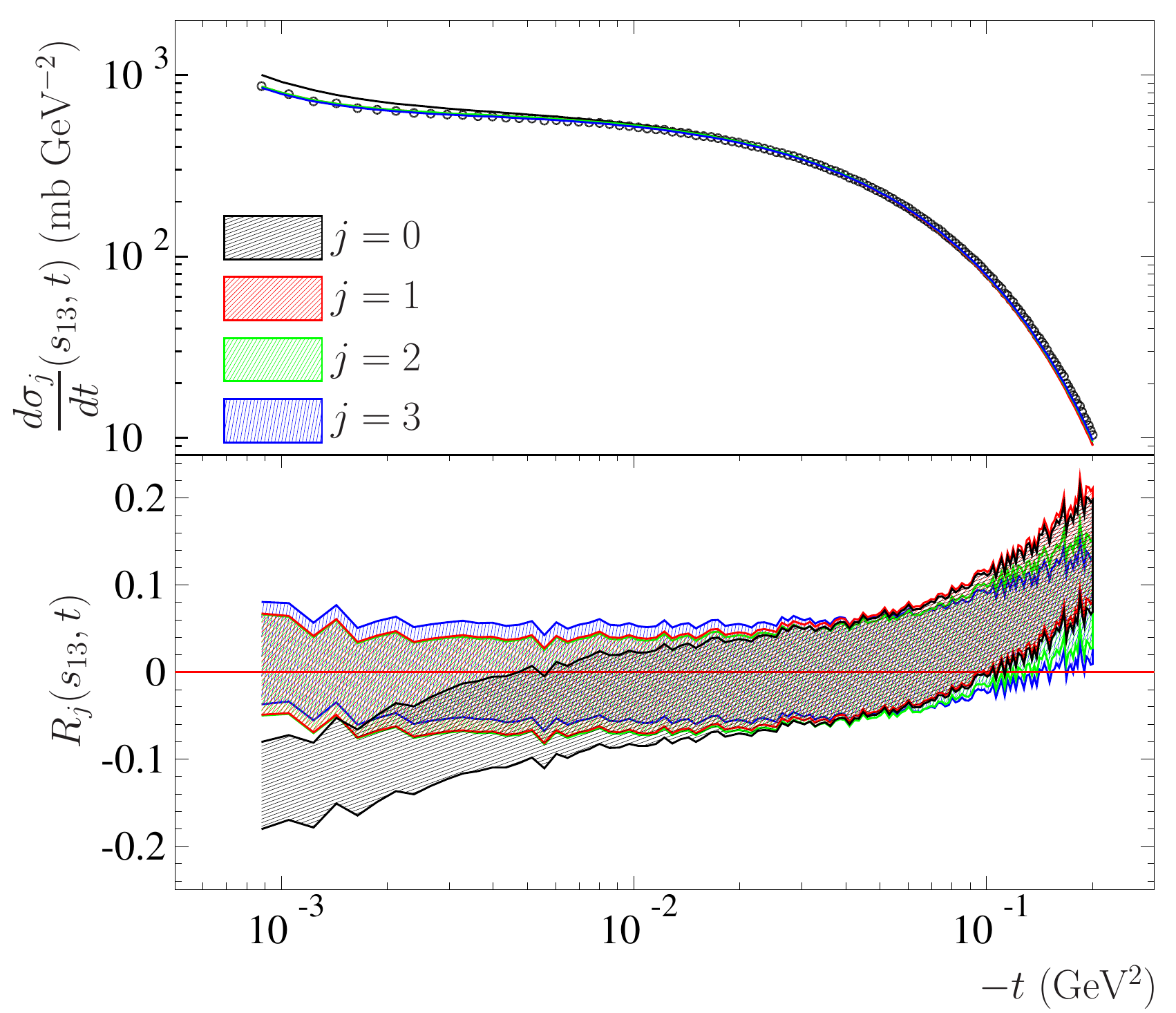}
\caption{\label{fig:diff-xs13}(Upper) data on the differential cross section at $\sqrt{s}=13$~TeV and superimposed the predictions corresponding to the parametrizations of the nuclear amplitude given in Eq.~(\ref{B10}). (Lower) Residuals as defined in Eq.~(\ref{eq:residuals}).}	
\end{center}
	\end{figure}

We now show our results in Figs.~\ref{fig:diff-xs8} and~\ref{fig:diff-xs13} for the complete elastic differential cross section in the CNI region at $\sqrt{s}=8$ and 13~TeV covering the $t$ region up to $|t|_{\rm max} = 0.2\ {\rm GeV}^2$. It is obtained from the modified nuclear BP amplitude discussed above in Sec.~\ref{BP}. We emphasize that in Figs.~\ref{fig:diff-xs8} and~\ref{fig:diff-xs13}, all four theoretical curves are drawn using the energy  behavior of the parameter  from Eqs.~\eqref{B4}-\eqref{B8} obtained without changing any parameters from Ref.~\cite{Fagundes:2013}, and implementing the rotation proposal of Eqs.~\eqref{B10}. Hence, these are {\it predictions} for the absolute differential cross section at $\sqrt{s}=8$ and 13~TeV as well as for the total cross section $\sigma_{\rm tot}$, the elastic one $\sigma_{\rm el}$ and $\rho$ parameter, that are discussed later.

Figures~\ref{fig:diff-xs8} and~\ref{fig:diff-xs13} show differential cross sections data, model predictions $d\sigma_j/dt(s_{8,13},t)$ and the corresponding residuals $R_j(s,t)$, in the four cases labeled with $j=0,1,2,3$ and at $\sqrt{s}= 8$ and 13~TeV. The $j$th differential cross section is obtained as
\ba
\frac{d\sigma_j}{dt}(s,t) = \Big|\mathcal{A}_j(s,t) + \mathcal{A}^C(s,t)\Big|^2\,,
\label{CNI}
\ea
 where the nuclear $j$th and Coulomb amplitudes are given in Eqs.~(\ref{B10}) and~(\ref{eq:A^C}) respectively, and residuals for the two datasets at $s=s_8$~ and $s=s_{13}$,
\ba
\left\{ t_k,\, \frac{d\sigma}{dt}(s,t_k)_{\rm data},\,\delta\lq \frac{d\sigma}{dt}(s,t_k)_{\rm data}\rq \right\}_{k=1}^{M(s)}\,,
\label{eq:dataset}
\ea
consisting of $M(s_8)$ and $M(s_{13})$ points, are defined as
\ba
	R_j(s,t_k)=\frac{\frac{d\sigma}{dt}(s,t_k)_{\rm data}-\frac{d\sigma_j}{dt}(s,t_k)}{\frac{d\sigma_j}{dt}(s,t_k)}\,,\hspace{2mm}
	\begin{array}{rcl}
		s&=& s_8,s_{13}\,,\\
		j&=& 0,1,2,3\,,\\
		k&=& 1,\ldots,M(s)\,.\\
	\end{array}
	\label{eq:residuals}
\ea  
As one can see, the agreement with data is excellent for the nuclear amplitudes $\mathcal{A}_{1,2,3}(s,t)$ (red, green, blue areas and lines), whereas $\mathcal{A}_{0}$ (black area and line) is essentially ruled out. The corresponding residuals $R_{1,2,3}(s,t)$ both at $\sqrt{s}=8$ and 13~TeV are practically zero all the way up to $|t|_{\rm max} = 0.1$~GeV$^2$. 

\begin{figure}[h!]
	\includegraphics[width=\columnwidth]{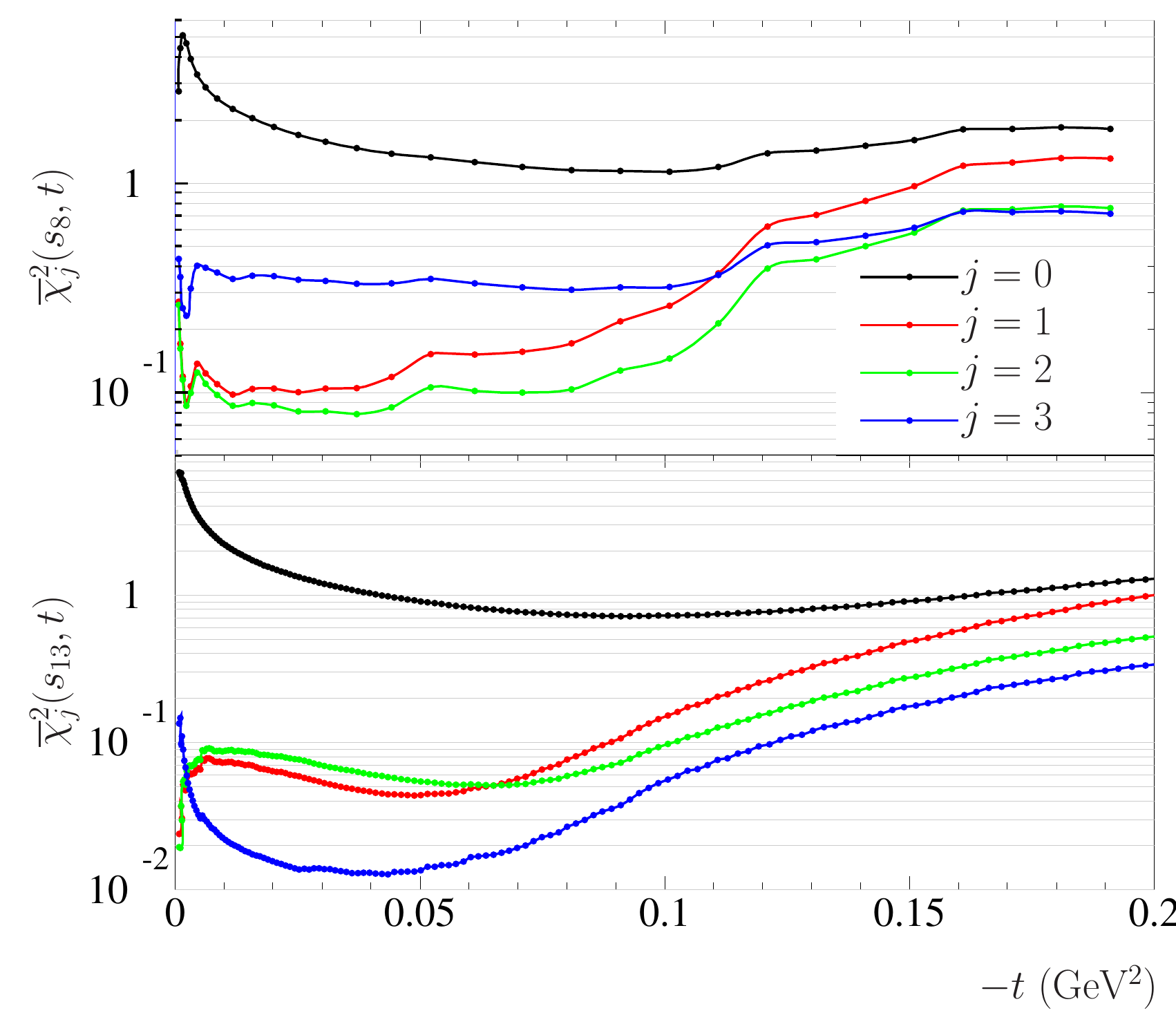}
	\caption{\label{fig:chi2}$\chi^2$ per degree of freedom in the four cases of Eq.~(\ref{B10}), labeled with $j=0,1,2,3$, and for the two sets of data at $\sqrt{s}=8$~TeV (top)
	and 13~TeV (bottom).}
\end{figure}

Figure~\ref{fig:chi2} shows, for the datasets of Eq.~(\ref{eq:dataset}), the $\chi^2_j(s,t)$ per degree of freedom 
\ba
\overline{\chi}_j^2(s,t_k)=\frac{1}{k}\sum_{l=1}^{k}\lt
\frac{\frac{d\sigma_j}{dt}(s,t_l)-\frac{d\sigma}{dt}(s,t_l)_{\rm data}}{\delta\left[\frac{d\sigma}{dt}(s,t_l)_{\rm data}\right]}\rt^2\,,
\no\ea
with $s=s_8,$ $s_{13}$ and $0<|t_k|<0.2$~GeV$^2$. The remarkably low value of $\overline{\chi}^2_{1,2,3}(s_{8,13},t)$, which is less than  0.3 for all $|t| \leq 0.1$~GeV$^2$, tells us that the nuclear amplitudes $\mathcal{A}_{1,2,3}(s,t)$ describe both $\sqrt{s}=8$ and 13~TeV data extremely well. We can then conclude that our model in the proposed analytic and crossing-symmetric version is very appropriate to describe present low-$|t|$ data, in the CNI region, and eventually predict future trends.
\begin{figure}[h!]
\begin{center}
\includegraphics[width=\columnwidth]{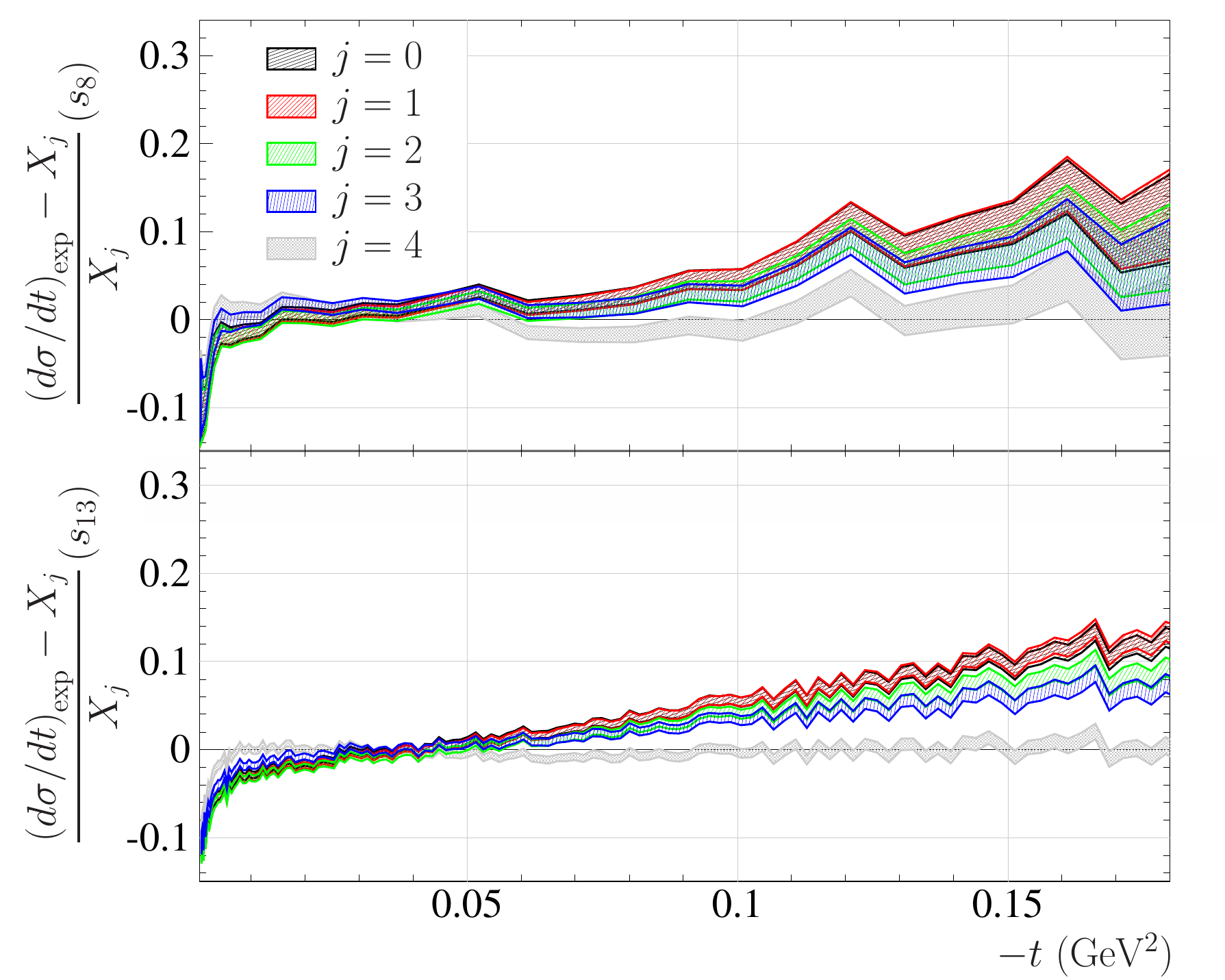}
\caption{\label{fig:no-interf}Residuals between data and Ref$_j(s,t)$, the noninterfering nuclear and Coulomb cross section defined in Eqs.~(\ref{eq:refj}) and (\ref{eq:ref4-13}), at $\sqrt{s}=8$~TeV (upper) and at $\sqrt{s}=13$~TeV (lower).}	\end{center}
	\end{figure}
Figure~\ref{fig:no-interf} shows the residuals 
\ba
\!\!\!\!\frac{\lt d\sigma/dt\rt_{\rm exp}-X_j}{X_j}(s)
\,,\hh j=0,1,2,3,4\,,\hh s=s_8,s_{13}\,,
\ea
between data $\lt d\sigma/dt\rt_{\rm exp}$ and the sum of Coulomb and nuclear cross sections with no interference, i.e., 
\ba
X_j(s,t)=
\Big|\mathcal{A}_j(s,t)\Big|^2 +\Big| \mathcal{A}^C(s,t)\Big|^2\,,
\label{eq:refj}
\ea
where the first four nuclear amplitudes, with $j=0,1,2,3$, are given in Eq.~(\ref{B10}), while the fifth one, with $j=4$, is obtained by using for the nuclear cross section the best exponential fits of Refs.~\cite{TOTEM:2016,TOTEM2:2017}. The expression at $\sqrt{s}=13$~TeV is given in Eq.~(\ref{eq:ref4-13}), while the one at $\sqrt{s}=8$ TeV reads
\ba
{\rm Ref}_4(s_8,t)&=&
527.1\,e^{\frac{19.39 \, t}{{\rm GeV}^{2}}}\,{\rm mb\,\,GeV}^{-2}
 +\Big| \mathcal{A}^C(s_{8},t)\Big|^2\,.
 \no\ea
Figure~\ref{fig:no-interf}, showing residuals between data and models without the CNI term, is the same as Figs.~16 and 17 of Ref.~\cite{TOTEM:2016} and Fig.~14 of Ref.~\cite{TOTEM2:2017}.
\section{Complex nuclear amplitude at $\sqrt{s}=8$ and 13~TeV \label{complex}}
As seen in the previous Secs.~\ref{differential} and~\ref{sec:noi}, nuclear amplitudes $\mathcal{A}_{1,2,3}(s,t)$ describe the differential cross section data in the CNI region very well. Here we shall exhibit the phases and both real and imaginary parts of all four of our nuclear amplitudes, in order to understand better qualitative differences between them and delineate further the problems associated with finding the correct value of the $\rho$ parameter.  

\begin{figure}[h!]\large
	\includegraphics[width=\columnwidth]{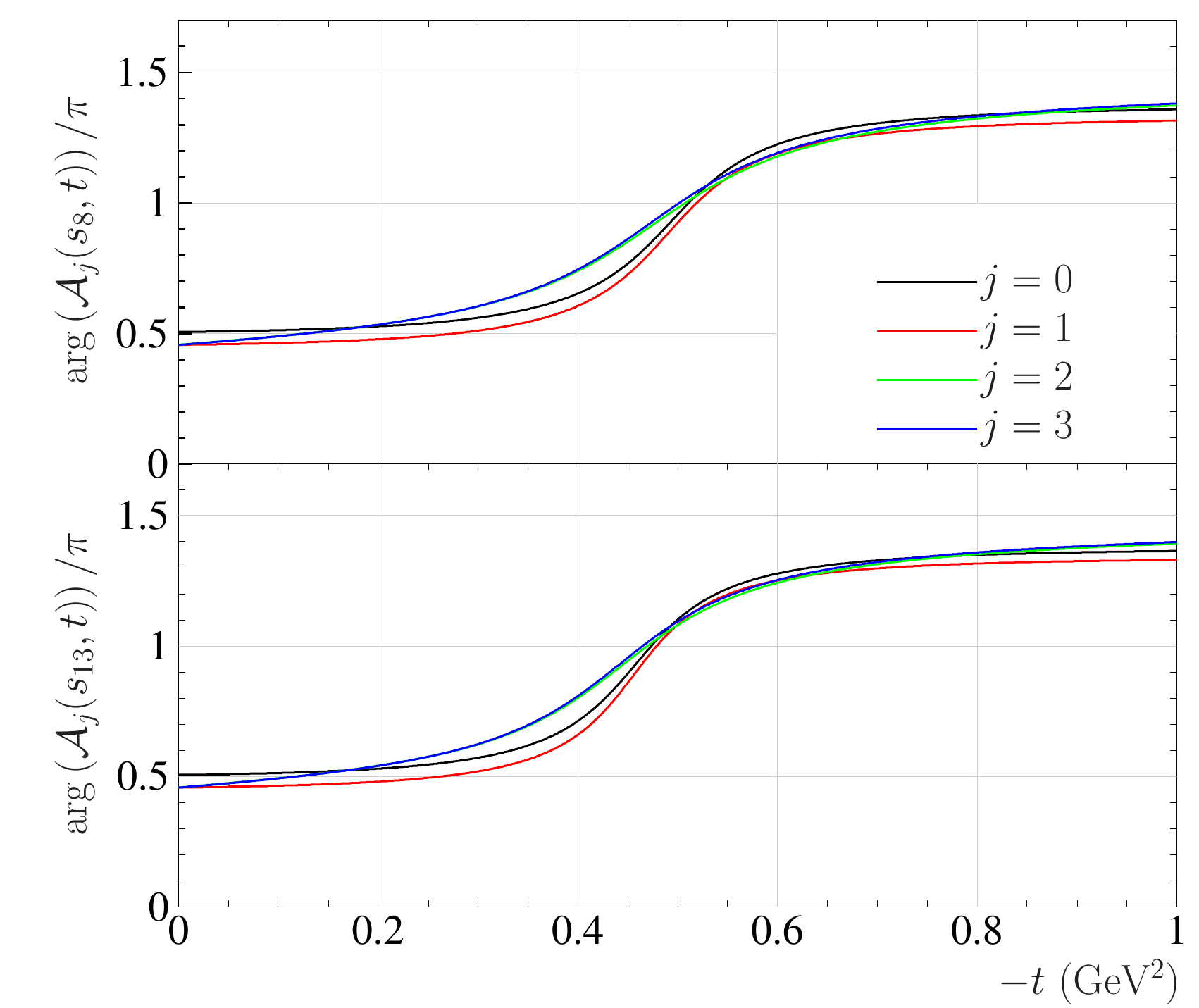}
	\caption{Phases of the four amplitudes $\mathcal{A}_j$, at $s=s_8$ (upper) and $s=s_{13}$ (lower), with $j=0$, (black) highest at $t=0$, $j=1$ (red) lowest at $t=0$, the curves related to the cases with $j=2,3$ (green and blue) are undistinguishable.}
\label{fig:phases}
\end{figure}

In Fig.~\ref{fig:phases} we show the phases predicted by the present model for all the four amplitudes previously described.

\begin{figure}[h!]
	\includegraphics[width=\columnwidth]{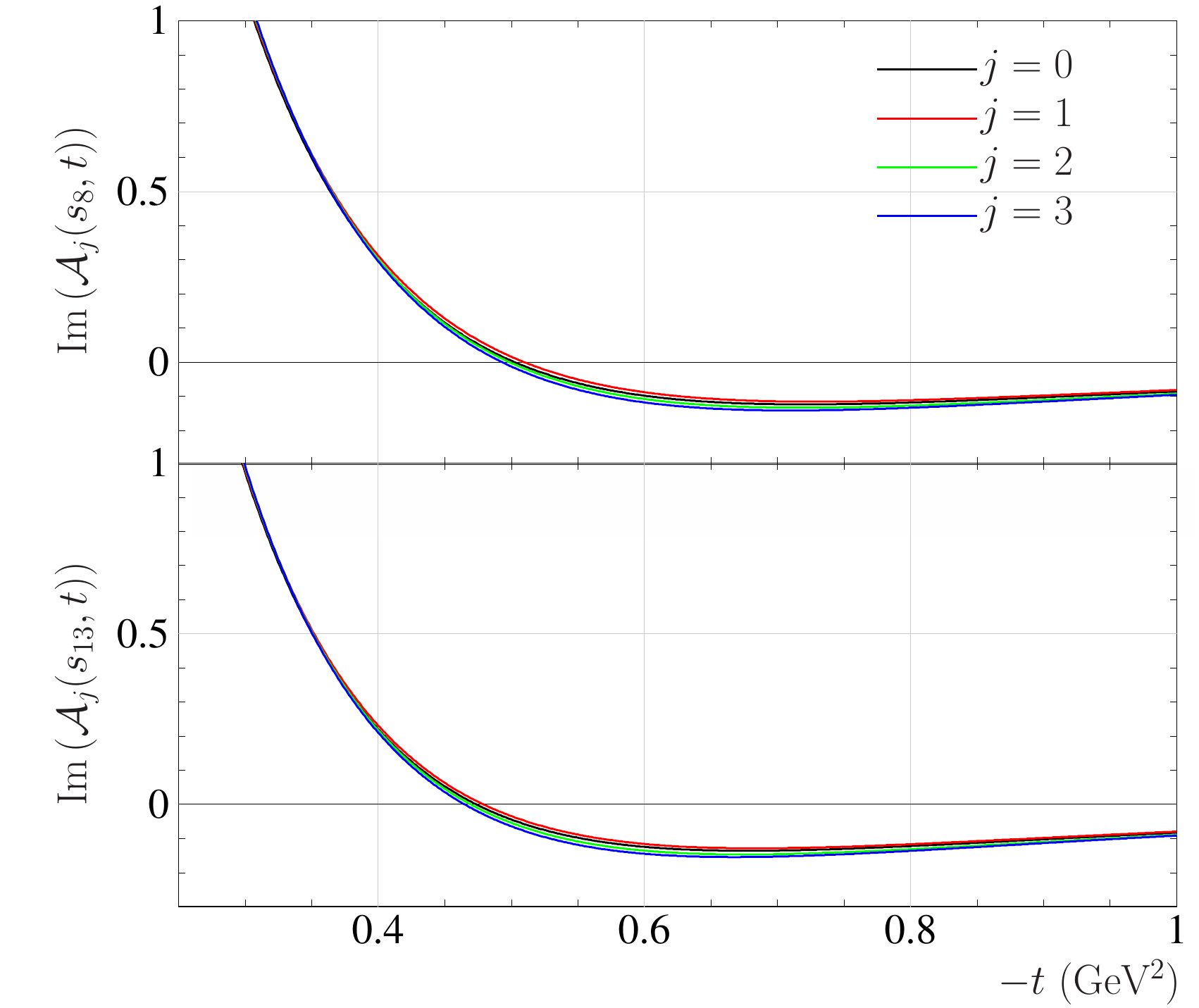}
\caption{Imaginary parts of the nuclear amplitudes at $s=s_8$ (upper) and $s=s_{13}$ (lower).}
\label{fig:imag}
\end{figure}

\begin{figure}[h!]
	\includegraphics[width= \columnwidth]{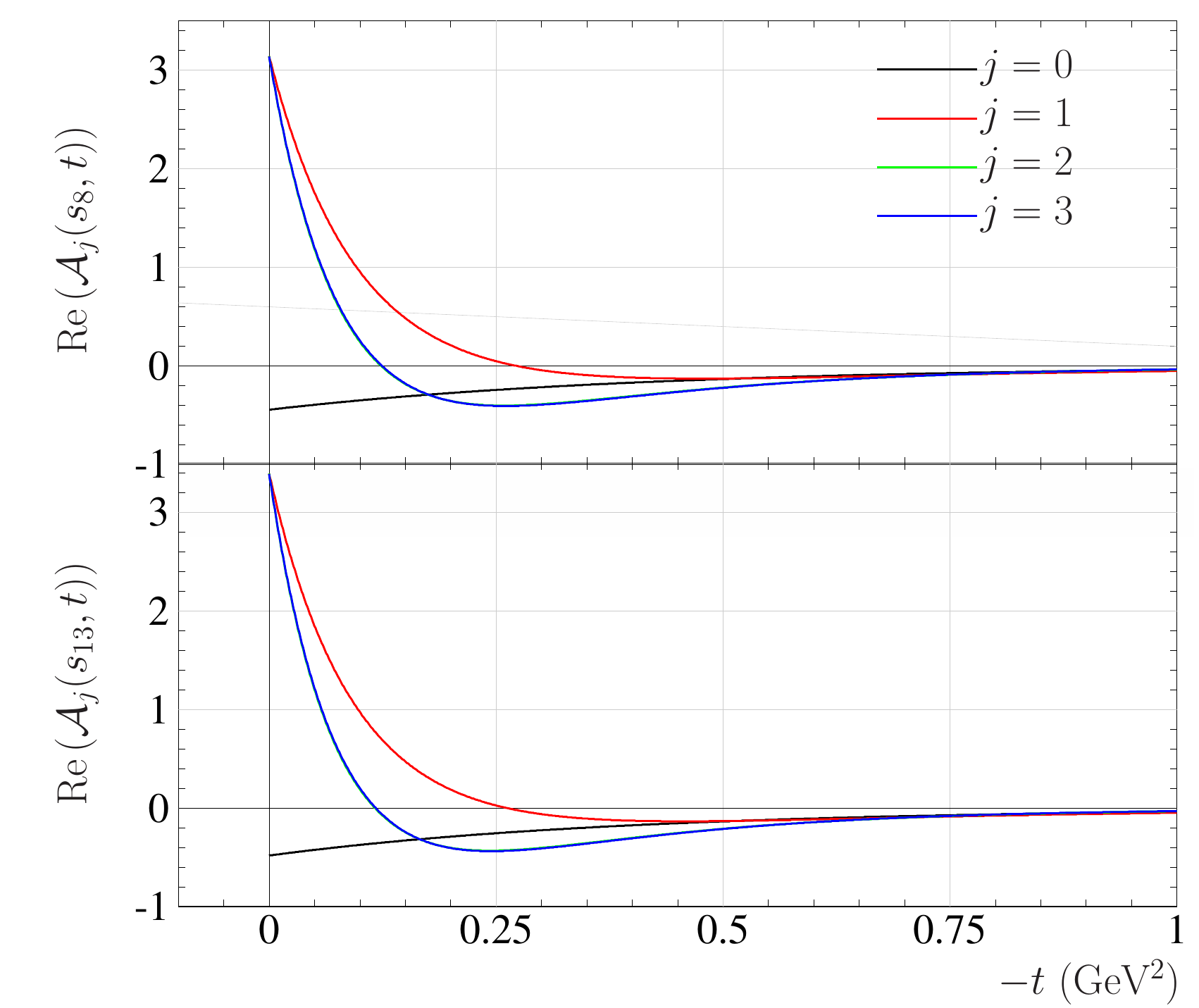}
\caption{Real parts of the four amplitudes $\mathcal{A}_j$, at $s=s_8$ (upper) and $s=s_{13}$ (lower), with $j=0$, (black) lowest at $t=0$, $j=1$ (red) highest at $t=0$, the curves related to the cases with $j=2,3$ (green and blue) are undistinguishable.}
\label{fig:real}
\end{figure}

Figures~\ref{fig:imag} and~\ref{fig:real} show imaginary and real parts of the amplitudes. From Fig.~\ref{fig:imag}, we see that the imaginary part is basically  unaffected by the crossing implementation. On the contrary, as it is shown in Fig.~\ref{fig:real}, the real part of the amplitude with no rotation $\mathcal{A}_0$ is clearly different from the other three real parts, that still are quite similar. In addition, and most importantly, Fig.~\ref{fig:real}  shows  that when rotated, the real part of the amplitude develops a zero  whereas the original unrotated expression for the real part is  negative throughout the entire forward peak region, up to $|t| \simeq 1$ GeV$^2$. Furthermore, even for the three rotated amplitudes, there are  differences in the position of the zero of the real part. In particular, we notice that:
\begin{itemize}
\item[(i)] in the cases $j=2$ and $j=3$, the two amplitudes are indistinguishable, with the real part developing a zero around $|t|=0.12$ GeV$^2$; 
\item[(ii)] for the $j=1$ case, with no rotation of slopes $B(s)$ ad $D(s)$, the real part, and hence the $\rho$ parameter, as a function of $t$, changes sign around $|t|=0.25$ GeV$^2$, already outside the CNI. 
\end{itemize}
We now turn to an estimate of the $\rho$ parameter in the small $|t|$ region. We will have to distinguish the case when the zero of the real part of the amplitude is within the CNI region, where the TOTEM experiment assumes a constant value~\cite{TOTEM2:2017}. By considering different physics assumptions and mathematical modeling, TOTEM has extracted by the same set of data on the proton-proton differential cross section, two values for the $\rho$ parameter, i.e.~\cite{TOTEM2:2017} 
\ba
\begin{array}{rcl}
\rho_{\rm T1}&=&0.09\pm 0.01\,,\\
\rho_{\rm T2}&=&0.10\pm 0.01\,.\\
\end{array}
\label{eq:totem-results}
\ea

\begin{figure}[h!]
\includegraphics[width=\columnwidth]{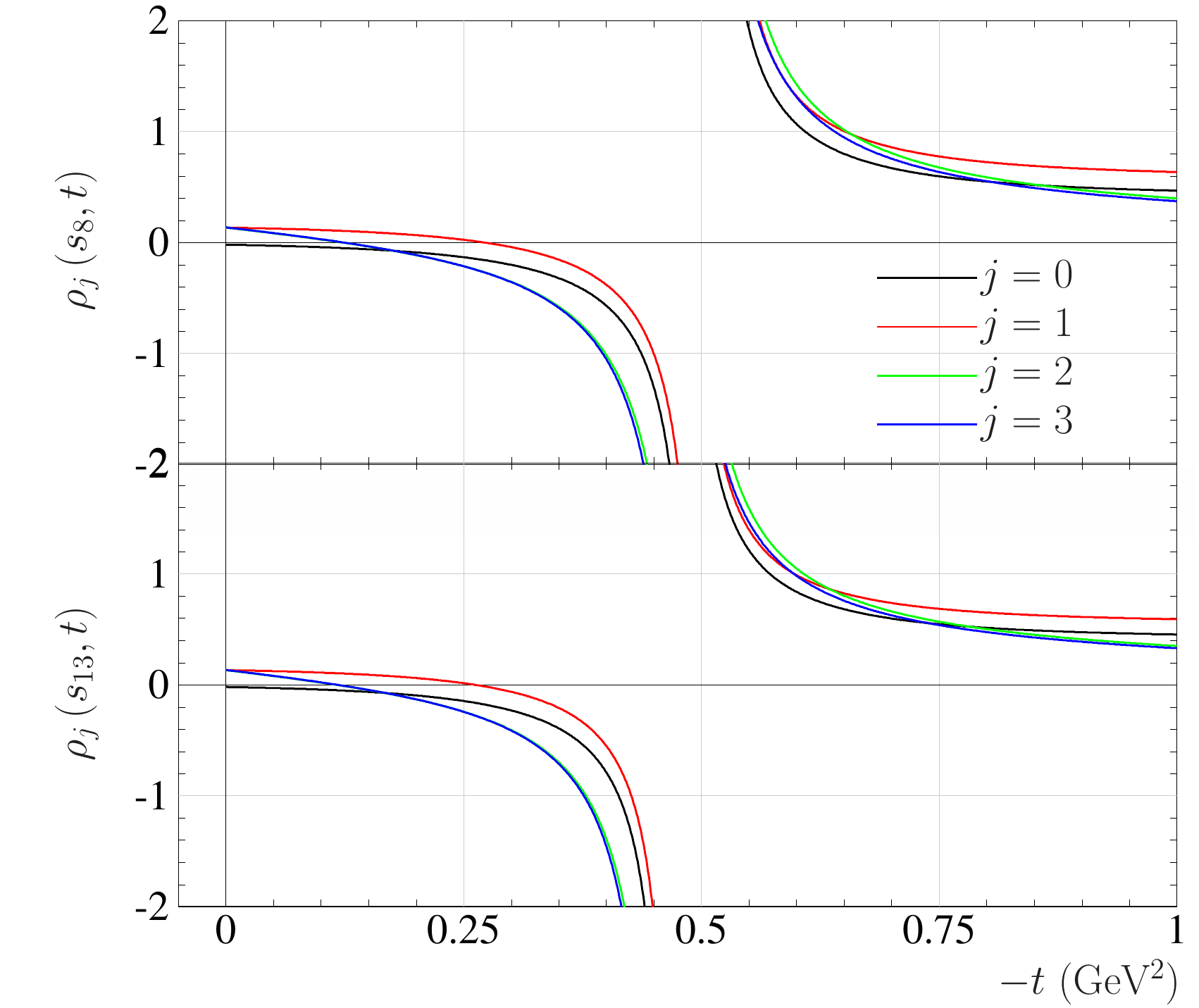}
\caption{
The $\rho$ values at $s=s_8$ (upper) and $s=s_{13}$ (lower), the curves of the cases with $j=0,1$ (black and red) are the lowest and highest ones at $t=0$, while those with $j=2,3$ are undistinguishable.}
\label{fig:rho}
\end{figure}

In Fig.~\ref{fig:rho} the $\rho$ parameters
\ba
	\rho_j(s,t)=\frac{\re\lt\mathcal{A}_{j}(s,t)\rt}{\im\lt\mathcal{A}_{j}(s,t)\rt}\,,
\ea
for the four amplitudes with $j=0,1,2,3$, at $s=s_8$ and $s=s_{13}$, are plotted as a function of $t$ in the momentum transfer region $0\le |t|\le 1$~GeV$^2$.
\section{Averages over momentum transfer and comparison with TOTEM results\label{meanrho}}
As discussed above and explicitly seen in Fig.~\ref{fig:real} for models $j=1,2,3$, the real part of the amplitude has a zero near or inside the CNI region. Thus, it is useful to define a mean value of $\rho$ that could be used to compare to the constant $\rho$'s used by TOTEM to analyze their data in the CNI region.

\begin{figure}[h!]
		\includegraphics[width=\columnwidth]{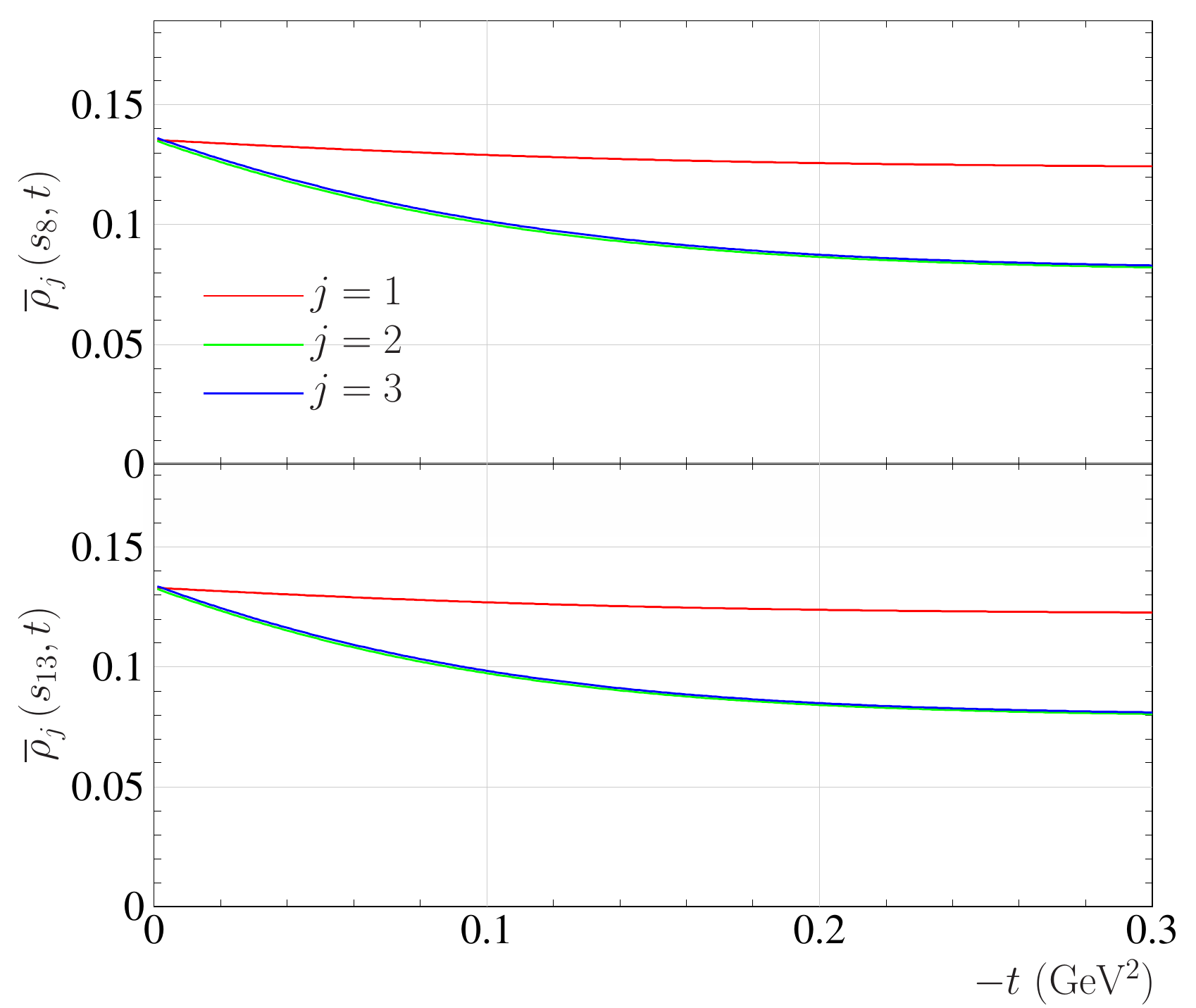}		
		\caption{\label{fig:rhomean}Mean values of $\rho_j(s,t)$ at $s=s_8$ (upper) and $s=s_{13}$ (lower), the curve of the case $j=1$ (red) is the highest for all $|t|$ values considered, the curves related to the other cases, $j=2,3$ (green and blue), are undistinguishable.}
\end{figure}

For this purpose, we consider a range of momentum transfer through  the region covered by the TOTEM data at $\sqrt{s}=8$ and $13$~TeV and average the model predictions for $\rho_j(s,t)$ in such ranges; namely, we define 
\ba
\overline{\rho}_j(s,t)=\frac{\ds\int_{-t}^0\rho_j(s,t')
\frac{d\sigma_j}{dt}(s,t')dt'}{\ds\int_{-t}^0 
\frac{d\sigma_j}{dt}(s,t')dt'}\,,\hspace{2mm} j=0,1,2,3\,,
\label{rhobar}
\ea
that are mean values depending on the $t$ interval ($0\le |t'| \le |t| $)--over the CNI region--as chosen by TOTEM. The obtained $\overline{\rho}_j(s,t)$, for the most relevant three cases $j=1,2,3$ are shown in Fig.~\ref{fig:rhomean}.

It is satisfactory that
\ba
\overline{\rho}_{2,3}(s_{13}, -0.15\ {\rm GeV}^2)
\simeq 0.09\,,
\nonumber
\ea
i.e., in the cases $j=2,3$, the mean value of $\rho$ in the momentum transfer region $0\le|t|\le 0.15$ GeV$^2$, at $\sqrt{s}=13$~TeV, is remarkably close to the value found by TOTEM for $\rho$, namely $\rho_{\rm T1}$ and $\rho_{\rm T2}$ of Eq.~\eqref{eq:totem-results}, {\it assumed to be constant quantities in this $t$ interval}.
\begin{table}[htb]
\caption{\label{tab:compendiumrho}%
Values for the $\rho$ parameter at zero transfer momentum and average over the interval $[-t,0]$, at $\sqrt{s}=8$ and 13~TeV from the rotated empirical model of Eq.~(\ref{B10}) and  Ref.~\cite{Fagundes:2013}. The corresponding TOTEM experimental values are also included together with the bibliographic references.}
\def\arraystretch{1.25}
\begin{tabular}{c|c|c|c|c|c}
\hline
$\sqrt{s}$& 
\multirow{2}{*}{$j$} & 
\multirow{2}{*}{$\rho_{j}(s,0)$} & 
\multirow{2}{*}{$\overline{\rho}_{j}(s,t)$} & 
$-t$ & 
\multirow{2}{*}{$\rho_{\rm TOTEM}^{\rm exp}$} \\
(Tev) & & & & (GeV$^2$) & \\
\hline
\hline
\multirow{3}{*}{8}	& 1 & 0.1352 & 0.1256 
& \multirow{3}{*}{0.2}
& \multirow{3}{*}{$0.12\pm0.03$~\cite{TOTEM:2016}}\\ 
\cline{2-4}
    					& 2 & 0.1352 & 0.0865 & & \\ 
\cline{2-4}
    					& 3 & 0.1365 & 0.0874 & & \\ 
\hline\hline
\multirow{3}{*}{13}	& 1 & 0.1330 & 0.1285 
& \multirow{3}{*}{0.07}
& \multirow{3}{*}{
\begin{minipage}{31mm}
$0.09\pm0.01$ ($N_b=1$)~\cite{TOTEM2:2017}\\
$0.09\pm0.01$ ($N_b=2$)~\cite{TOTEM2:2017}\\
$0.10\pm0.01$ ($N_b=3$)~\cite{TOTEM2:2017}\\
\end{minipage}
}\\ 
\cline{2-4}
    					& 2 & 0.1330 & 0.1050 & & \\ 
\cline{2-4}
    					& 3 & 0.1341 & 0.1062 & & \\ 
\cline{1-6}
\multirow{3}{*}{13}	& 1 & 0.1330 & 0.1247 
& \multirow{3}{*}{0.15}
&
\multirow{3}{*}{
\begin{minipage}{31mm}
$0.09\pm0.01$ ($N_b=2$)~\cite{TOTEM2:2017}\\
$0.10\pm0.01$ ($N_b=3$)~\cite{TOTEM2:2017}\\
\end{minipage}
}
 \\ 
\cline{2-4}
    					& 2 & 0.1330 & 0.0877 & & \\ 
\cline{2-4}
    					& 3 & 0.1341 & 0.0885 & & \\ 
\hline
\end{tabular}
\end{table}
In Tables~\ref{tab:compendiumrho} and~\ref{tab:compendium}, we report the values obtained for the quantities of interest in this paper, i.e.: the $\rho$ parameter at $t=0$, its average value in the $t$ interval corresponding to the one investigated by TOTEM, and the total and elastic cross section at the two LHC energies. 

A more comprehensive analysis of the model, tuned for a larger $t$ interval covering the bump-dip region ($|t| \leq 1$ GeV$^2$) is left for future work after definitive TOTEM results become available with proper overall normalization over the entire momentum transfer interval. 
\begin{table}[h]
\caption{\label{tab:compendium}%
Values of $\sigma_{\rm tot}^{j}$, $\sigma_{\rm el}^{j}$ in the four cases $j=0,1,2,3$ at $\sqrt{s}=8$ and $13$~TeV from  the rotated empirical model of Eq.~(\ref{B10}) and Ref.~\cite{Fagundes:2013}. We also include  TOTEM experimental values together with the bibliographic references.} 
\def\arraystretch{1.25}
\begin{tabular}{c|c|c|c|c|c}
\hline
$\sqrt{s}$& 
\multirow{2}{*}{$j$} & 
$\sigma_{\rm tot}^j$ & 
$\sigma_{\rm tot}^{\rm exp}$ & 
$\sigma_{\rm el}^j$ & 
$\sigma_{\rm el}^{\rm exp}$ \\
(Tev) & & (mb) & (mb) & (mb) & (mb) \\
\hline
\hline
\multirow{4}{*}{8}	& 0 & 103.44
    & 
	& 26.82
	& \multirow{4}{*}{$27.1\pm1.4$~\cite{Stark:2017}}\\ 
\cline{2-3}\cline{5-5}
    					& 1 & 102.68 
    					&$103.2\pm2.3 $~\cite{TOTEM:2016}  & 26.85 & \\ 
\cline{2-3}\cline{5-5}
    					& 2 & 102.68 
    					&$102.9\pm2.3 $~\cite{TOTEM:2016}  & 27.06 & \\ 
\cline{2-3}\cline{5-5}
    					& 3 & 101.64 &  & 26.85 & \\ 
\hline\hline
\multirow{4}{*}{13}	& 0 & 113.66
	& \multirow{4}{*}{$110.6\pm3.4 $~\cite{TOTEM1:2017}}
	& 30.76
	& \multirow{4}{*}{$31.0\pm1.7$~\cite{TOTEM1:2017}}\\ 
\cline{2-3}\cline{5-5}
    					& 1 & 112.87 &  & 30.81 & \\ 
\cline{2-3}\cline{5-5}
    					& 2 & 112.87 &  & 31.03 & \\ 
\cline{2-3}\cline{5-5}
    					& 3 & 111.84 &  & 30.84 & \\ 
\hline
\end{tabular}
\end{table}
\section{Do TOTEM data require odderons?\label{odd}}
The precise TOTEM data in the CNI region at $\sqrt{s}=13$~TeV give us rather accurate values of the $\rho$ parameter~\cite{TOTEM2:2017} reported in Eq.~\eqref{eq:totem-results}. We have presented  in the previous sections an analysis  which confirms the measured low value of the $\rho$ parameter at $\sqrt{s}=13$~TeV, through an  empirical model  based on  analyticity, crossing symmetry and which satisfies known asymptotic theorems. As the existence of a zero in the CNI region, which  is found to be consistent with TOTEM results, was connected in Ref.~\cite{Martin:1997} to the asymptotic behavior of the total cross section, we shall now discuss this issue.

Specifically, we shall consider the possibility that the imaginary part of the scattering amplitude at $t=0$ may not saturate the Froissart bound. Let us compare the above with an assumed high energy total cross section increasing as some power of $L(s) = \ln(s/s_0)$ [see Eq.~\eqref{eq:Ls}]. Using the nomenclature from the soft-gluon resummation work reviewed in Ref.~\cite{Pancheri:2017}, we may write the forward elastic amplitude--using the rule discussed in Sec.~\ref{BP}--as follows. Considering the amplitude
\ba
\label{o2}
\mathcal{A}(s, 0) &=& i K \left[
\ln(s e^{-i\pi/2}/s_0)\right]^{1/p}\,, 
\ea
where $K$ is a positive constant, in the limit of large $s/s_0$, i.e., $L(s)\gg 1$, the imaginary and real parts of the amplitude read 
\ba
\begin{array}{rcl}
\im\lt \mathcal{A}(s,0)\rt &\simeq& K L^{1/p}(s)\,,\\
 \re \lt \mathcal{A}(s,0) \rt &\simeq& \displaystyle \
\frac{\pi}{2 p }\, K L^{1/p-1}(s)\,,\\
\end{array}
 \ea
 so that the total cross section and the $\rho$ parameter are
 \ba
 \label{o4}
\sigma_{tot}(s) = K L^{1/p}(s)\,,\hh \rho(s) = \frac{\pi}{2p L(s)}\,.
\ea
In such a model the parameter $p$, varying in the interval $[1/2,1]$, describes the level of saturation of the Martin-Froissart bound. The limit is reached for $p=1/2$ leading to the well-known Khuri-Kinoshita bound $\rho = \pi/L$~\cite{khuri}.\\
 
Of course, care should be exercised in applying such expressions and comparing them to data due to finite corrections as well as due to the occurrence of a zero in the real part of the nuclear amplitude in the near-forward CNI region, as it is only after averages over the $t$ intervals are taken that we obtain $\bar{\rho} \simeq (0.09\div0.1)$, in agreement with the TOTEM values.\\
Just for illustration, consider our analysis at 13\ TeV. Our nuclear amplitudes are chosen to saturate the Froissart bound (i.e., $p=1/2$ in the notation above). Thus, asymptotically, $\rho$ should have the KK value. Numerically, at 13\ TeV,  $\rho_{KK}= 0.166$. On the other hand, after finite corrections, this value is reduced to $\hat{\rho} \approx 0.15$ [see, Eq.~\eqref{eq:rhohat}]. Consideration of our complete nuclear amplitude (i.e., including contributions also from the $\sqrt{C}$ term), reduces it further to our actual value of  $\rho \simeq 0.133$ (see Table 1). 

Using Eq.~\eqref{o4}, this would be obtained for an effective $p\approx 0.62$ (modulo further finite corrections, if any). And for this value of $p$, $\sigma_{\rm tot}(s)\sim L^{1.61}(s)$, a slower rate than $\sigma_{\rm tot}(s)\sim L^2(s)$.\\
A similar trend is also reflected in other  phenomenological analyses, such as in the soft-gluon resummation model of Ref.~\cite{Corsetti:1996}, as shown in Ref.~\cite{Grau:2009qx}. Another example can be found in Table~13 of Block's review of Ref.~\cite{Block:2006}, $\sigma_{\rm tot}$ is parametrized with the leading term $c_2 L^2(s)$, where $c_2 = + 0.275$ mb is a (small) positive coefficient, accompanied by a nonleading term $c_1 L(s)$, where $c_1 = -1.3$ mb is instead a large negative coefficient. While TOTEM has not released a total cross section expression using a power series in $L(s)$, the following parametrization of the LHC elastic cross section data--attributed to Compact Muon Solenoid--can be found in the TOTEM report of Ref.~\cite{TOTEM:2011}
\ba
\sigma_{\rm el}(s) = \big[0.130\ L^2(s) - 1.5\ L(s) + 11.4\big]\, {\rm mb}\,.\no
\ea
The above expression suggests that indeed the proton-proton total, elastic and inelastic cross sections might be increasing at a rate lower than $L^2(s)$, as $s$ increases. 

The modified crossing symmetric BP near-forward nuclear amplitude discussed at length in the previous sections is also anchored upon two leading terms, $\sqrt{A(s)}$ and $\sqrt{C(s)}$, with the dominant $L^2(s)$ term in $A(s)$ with a positive coefficient followed by a nonleading term with a negative coefficient  [$- 3.8 L(s) $ in Eq.~(\ref{B4})] and by next $\sqrt{C(s)}$ term, as seen from Eq.~(\ref{B5}). The success of this model in describing TOTEM data in the CNI region also suggests a similar pattern without on the other hand invoking a $C = -1$ odderon contribution. Theoretical QCD models for $C = -1$ three gluon color singlet Regge trajectories obtain a very low intercept, even lower than the $\omega$ trajectory~\cite{Estrada:2006}. Also, at HERA, the H1 group~\cite{H1} has ruled out an odderon Regge intercept above $0.7$. Thus, the prognosis for QCD odderons with intercepts equal to or greater than 1 seems exceedingly dim. The odderon hypothesis may also have difficulties with unitarity and the black disk limit as recent QCD model calculations~\cite{Khoze:2018} show a contradiction between unitarity and a maximal odderon~\cite{Martynov:2017zjz}, namely the one which would give a contribution to the total cross section rising  $\sim \ln^2(s)$ . Also, a calculation in the color glass condensate model~\cite{Gotsman:2018buo} estimates the contribution of the $C$-odd  amplitude to $\rho$ to be of order 1\%, namely $\Delta \rho_{\rm odderon}\sim (1\,{\rm mb})/\sigma_{\rm tot}\lesssim 1\%$.

Further evidence that TOTEM results  do not force the existence of a $C=-1$ contribution in the CNI region can be found in a recent paper by Jenkovszky and co-workers~\cite{Jenkovszky:2018-2}. It is to be hoped that further TOTEM differential cross section data (not just total cross section data) covering a larger $t$ interval would help resolve this issue.   
\section{Conclusions \label{conc}}
Modified BP nuclear amplitudes, called models $j=2,3$ in the text, appear to describe the CNI data rather well thus allowing us to draw the following conclusions.
\begin{enumerate}
\item At LHC energies, the real part of the nuclear amplitude vanishes at a momentum transfer value $t\simeq-0.12$~GeV$^2$.
\item As TOTEM CNI data at $\sqrt{s}=8$~TeV~\cite{TOTEM:2016} were analyzed over the interval $6\times 10^{-4}\leq |t| \leq 0.19\ {\rm GeV}^2$, the zero in the real part of the nuclear amplitude was within this interval. This might explain the choice of a peripheral phase with double zeroes in both the real and the imaginary parts (see solution KL/peripheral in Table~5 of Ref.~\cite{TOTEM:2016}), which, however, violates analyticity and positivity. Such a peripheral phase solution can be ruled out on rigorous grounds. An analysis of the 8 TeV data, based on a simple Regge  model~\cite{Fagundes:2016rpx}, argued that the nonexponential behavior at low $|t|$ could be explained by the existence of a threshold  singularity~\cite{CohenTannoudji:1972gd,Jenkovszky:2018},  required by $t-$channel unitarity. On the other hand,  the empirical model discussed here with a single zero in both the real and the 
imaginary parts is able to adequately describe the CNI data at $\sqrt{s}=8$~TeV (as well as that $\sqrt{s}=13$~TeV).
\item For LHC energies, the zero in the real part lies in the CNI region, thus rendering a precise determination of $\rho$ parameter rather problematic. For example, TOTEM analyzed its $\sqrt{s}=13$~TeV data assuming that
\ba
\label{C1}
\rho(s,t) = \frac{\re \lt \mathcal{A}(s,t)\rt}{\im \lt \mathcal{A}(s,t)\rt} \equiv \rho(s)\,,
\ea
i.e., that the $\rho$ parameter is a constant in $t$ over the region $0\le |t| \le 0.15$ GeV$^2$, whereas we expect $\rho(s, t\simeq- 0.12\ {\rm GeV}^2) = 0$. To make a comparison, we defined a mean value, see Sec.~\ref{meanrho} etc. over this interval and found good agreement with the TOTEM value.
\item The need of a QCD odderon contribution to explain the TOTEM data in the CNI region is not compelling, in particular for what concerns the energy rise of the total cross section and the nonsaturation of the Froissart bound at present LHC energies. Thus, model-independent analyses of the entire $t$ region covered by future TOTEM data are likely to clarify this important issue.
\item Both elastic and total cross sections appear to be approaching their asymptotic limits from below. For example     
\ba
\frac{\sigma_{\rm el}}{\sigma_{\rm tot}}\lq{\rm LHC \,\,energies}\rq < \frac{1}{3}
<\frac{\sigma_{\rm el}}{\sigma_{\rm tot}}\lq{\rm black\, \,disk}\rq\equiv \frac{1}{2}
\,.
\nonumber\ea
\item Also, the values for the $\rho$ parameter obtained by TOTEM at $\sqrt{s}=13$ TeV, $\rho_{\rm T1}$ and $\rho_{\rm T2}$ of Eq.~\eqref{eq:totem-results}, are much lower than the Khuri-Kinoshita bound
\ba
\frac{\pi}{L(s_{13})} \simeq\ 0.165\,,\nonumber
\ea
having $L(s_{13})\simeq 18.95$.
\item Both items 5 and 6 above suggest that cross sections very likely rise less fast than $L^2(s)$~\cite{Pancheri:2017}. Clearly, further data are required to settle this crucial issue. 	
\end{enumerate}\vspace{4mm}

\section{Acknowledgements}
Y. S. would like to thank the Department of Physics and Geology at the University of Perugia, Perugia, Italy for its hospitality. We thank Professor Earle Lomon, from CTP at MIT, for invaluable advice and criticism during the preparation of this paper.

\end{document}